\documentclass[journal=ancac3,manuscript=article,layout=twocolumn]{achemso}
\usepackage[version=3]{mhchem}
\usepackage{siunitx}
\usepackage{bbding}
\usepackage{soul}
\usepackage{verbatim}
\usepackage{scalerel}
\usepackage{multicol}
\usepackage{xr}
\usepackage{cleveref}
\usepackage{abstract}
\usepackage[fontsize=11pt]{fontsize}
\usepackage{chemformula} 
\usepackage[T1]{fontenc} 
\usepackage{etoolbox}

\let\oldmaketitle\maketitle
\let\maketitle\relax

\title[An \textsf{achemso} demo]{Long-range three-dimensional tracking of nanoparticles using interferometric scattering (iSCAT) microscopy}

\makeatletter
\renewcommand*{\acs@author@fnsymbol@symbol}[1]{
    \ifcase #1 \text{§}\or 
    1\or
    2\or
    3\or
    *\or     
    \fi
}
        
\renewcommand*\acs@contact@details{
    {\sffamily \text{§}\,E-mail: \acs@email@list }%
    \acs@number@list
}           

\patchcmd{\acs@address@list@auxii}
{\acs@author@fnsymbol{\acs@affil@marker@cnt}}
{\textsuperscript{\acs@author@fnsymbol{\acs@affil@marker@cnt}}}
{}{}

\patchcmd{\acs@address@list@auxii}
{{\acs@author@fnsymbol{\acs@affil@marker@cnt}\@nameuse{@altaffil@\@roman\@tempcnta}\par}}
{{\textsuperscript{\acs@author@fnsymbol{\acs@affil@marker@cnt}}\@nameuse{@altaffil@\@roman\@tempcnta}\par}}
{}{}        

\makeatother    
\makeatletter
\newcommand\myabstract[1]{%
  \if@twocolumn
    \@restonecoltrue\onecolumn
    \twocolumn[\begin{@twocolumnfalse}\oldmaketitle\begin{abstract}#1\end{abstract}\end{@twocolumnfalse}]%
  \else
    \oldmaketitle
    \begin{abstract}#1\end{abstract}
  \fi
}
\makeatother

\title[An \textsf{achemso} demo]{Long-range three-dimensional tracking of nanoparticles using interferometric scattering (iSCAT) microscopy}

\author{Kiarash Kasaian}
\affiliation[MPL]{Max Planck Institute for the Science of Light, 91058 Erlangen, Germany.}
\alsoaffiliation{Max-Planck-Zentrum f{\"u}r Physik und Medizin, 91058 Erlangen, Germany.}
\alsoaffiliation{Department of Physics, Friedrich-Alexander-Universit{\"a}t Erlangen-N{\"u}rnberg, 91058 Erlangen, Germany.}
\altaffiliation{These authors contributed equally.}

\author{Mahdi Mazaheri}
\affiliation[MPL]{Max Planck Institute for the Science of Light, 91058 Erlangen, Germany.}
\alsoaffiliation{Max-Planck-Zentrum f{\"u}r Physik und Medizin, 91058 Erlangen, Germany.}
\alsoaffiliation{Department of Physics, Friedrich-Alexander-Universit{\"a}t Erlangen-N{\"u}rnberg, 91058 Erlangen, Germany.}
\altaffiliation{These authors contributed equally.}

\author{Vahid Sandoghdar}
\affiliation[MPL]{Max Planck Institute for the Science of Light, 91058 Erlangen, Germany.}
\alsoaffiliation{Max-Planck-Zentrum f{\"u}r Physik und Medizin, 91058 Erlangen, Germany.}
\alsoaffiliation{Department of Physics, Friedrich-Alexander-Universit{\"a}t Erlangen-N{\"u}rnberg, 91058 Erlangen, Germany.}
\email{vahid.sandoghdar@mpl.mpg.de}


\begin{document}

\myabstract{
  Tracking nanoparticle movement is highly desirable in many scientific areas, and various imaging methods have been employed to achieve this goal. Interferometric scattering (iSCAT) microscopy has been particularly successful in combining very high spatial and temporal resolution for tracking small nanoparticles in all three dimensions. However, previous works have been limited to an axial range of only a few hundred nanometers. Here, we present a robust and efficient strategy for localizing nanoparticles recorded in high-speed iSCAT videos in three dimensions over tens of micrometers. We showcase the performance of our algorithm by tracking gold nanoparticles as small as \SI{10}{\nano\metre} diffusing in water while maintaining \SI{5}{\micro\second} temporal resolution and nanometer axial localization precision. Our results hold promise for applications in cell biology and material science, where the three-dimensional motion of nanoparticles in complex media is of interest. }


\section{Introduction}   
   
Single particle tracking (SPT) is a powerful technique for investigating the dynamic interaction of individual nanoparticles with heterogeneous environments \cite{Saxton-1997,Manzo-2015}.  The key step in SPT is to image an isolated nano-object onto a well-defined intensity distribution, namely the point-spread function (PSF) of the optical system in use. By fitting a known theoretical or experimental model to the PSF, one can pinpoint the particle location in each video frame and establish its trajectory over time. It follows that the localization precision in each frame is dictated by the signal-to-noise ratio (SNR) of the PSF over its background, whereby the signal, background and noise levels depend on various imaging modalities and sample conditions \cite{Nguyen-2023,mortensen-2010}. The available SNR puts a fundamental limit on the size of a nano-object and the speed with which it can be tracked. As an example, large signals from particles such as a micrometer-sized bead used in optical tweezer experiments can yield \text\AA ngstrom localization precision within 0.1\,s \cite{huhle-2015-tweezer}. Over the past three decades, SPT has been extensively applied to studies of diffusion and transport in very different contexts, spanning cell biology and biophysics \cite{VonDiezmann2017,Reina2018} to material science \cite{Weeks2000} and statistical physics \cite{Metzler2014,Mazaheri-2020,Beckwith2021}. 

The PSF in conventional microscopy techniques such as fluorescence and dark-field scattering is solely based on intensity and can usually be approximated by the profile of a Gaussian beam. Because in these methods the PSF is more extended in the third dimension, the axial localization precision is lower than in the lateral plane. Furthermore, it becomes increasingly difficult to track particles that move away from the imaging plane. As a result, the great majority of works have only recorded two-dimensional (2D) projections of the 3D particle trajectories. Many methods such as multi-focal plane imaging\cite{Wells-2010,Wang-2019,Louis-2020} and PSF engineering\cite{Moerner2010} have been applied to extend the axial range. A powerful alternative approach for performing high-precision axial tracking is to use interferometric microscopy \cite{Taylor-2019,martin-2022}. Interferometric measurements are particularly advantageous due to the use of phase information along the axial direction, which allows precise monitoring of the motion of a particle away from the imaging plane \cite{Verpillat2011}.

\section{Interferometric single-particle tracking (iSPT)}

Interferometric methods such as holography have been used to track particles in different arrangements, albeit mostly addressing relatively large objects \cite{martin-2022}. In the case of particles much smaller than the wavelength of light, the optical response is governed by Rayleigh scattering, and the method of choice is referred to as interferometric scattering microscopy (iSCAT) \cite{Lindfors-2004,Taylor-2019-nano}. This method exploits a homodyne detection scheme, where the scattered electric field from a nano-object interferes with the field of a reference beam. In the wide-field mode (see Figure \ref{setup_PSF_calibration}a), a nearly-collimated illumination is realized by focusing a light beam at the back focal plane of the microscope objective. In the most common form of iSCAT, the reference beam is constituted by reflection from the interface between the sample medium and the substrate supporting it. The detected iSCAT signal in this arrangement can be written as 
\begin{equation}   
  I_{det} \propto \lvert E_{ref}\rvert^2\,+\,2\lvert E_{ref}\rvert \lvert E_{sca}\rvert \cos{\phi}\,+\,\lvert E_{sca}\rvert^2\,,
   \label{eq:1}   
\end{equation}  
where the reference field $E_{ref} = r E_{inc}$ stems is reflected from the medium-glass interface, $E_{sca} = s E_{inc}$ represents the light scattered from the sample, and $\phi$ is the phase difference between $E_{ref}$ and $E_{sca}$. To account for potential variations in the illumination beam, we normalize the iSCAT images to $I_{ref}=\lvert E_{ref}\rvert^2$ and define the contrast $C$ as   
 
\begin{equation}   
  C = \frac{I_{det}-I_{ref}}{I_{ref}}=2\frac{\lvert s\rvert}{\lvert r\rvert} \cos{\phi} + \frac{\lvert s\rvert^2}{\lvert r\rvert^2}\,.   
   \label{eq:2}   
\end{equation}

Over the past two decades, many efforts have demonstrated the remarkable sensitivity of iSCAT for detection of nanoparticles down to single small proteins \cite{Piliarik-2014,Young-2018,Dahmardeh-2023}. As compared to fluorescence SPT, iSCAT has a nearly infinite photon budget because it suffers neither from saturation nor from photobleaching. This provides access to both high temporal resolution and long-term studies in tracking  \cite{Kukura-2009,Lin-2014,Spindler-2016,Taylor-2019}. Another decisive advantage of iSCAT is that its interferometric nature makes the signal highly sensitive to the axial position of the nanoparticle under study \cite{jacobsen_2007, Krishnan-2010, deWit-2018}.  However, the ambiguity resulting from the periodicity of the modulating traveling phase prevents one from determining the direction of travel over a range longer than about $\lambda/4$, where $\lambda$ is the wavelength of light in the medium of interest. To get around this problem, one can exploit the axial asymmetry of the spherical aberration about the focal plane. It, thus, follows that the radial cross section of the interferometric point-spread function (iPSF) contains information about the height of the particle above the cover glass \cite{Mahmoodabadi-2020}. 

In our previous efforts, we exploited the axial asymmetry caused by spherical aberration and use an unsupervised machine learning scheme with \textit{k}-means clustering to demonstrate an axial range of approximately \SI{300}{\nano\meter} \cite{Taylor-2019, Mahmoodabadi-2020}. However, extension of this approach to longer axial ranges presented challenges such as increased computational complexity and potential inaccuracies in clustering, as it relies on the silhouette values for determining the optimal number of clusters. In our current work, we establish a suitable estimator to assign the full lateral content of the experimental iPSF along a trajectory to the computed iPSFs specific to our iSCAT setup. To achieve this, we calibrate the imaging system carefully and establish a computational workflow to model the experimental iPSF.  

\subsection{The algorithm}         

\begin{figure*}[t!]   
  \includegraphics[width=\textwidth]{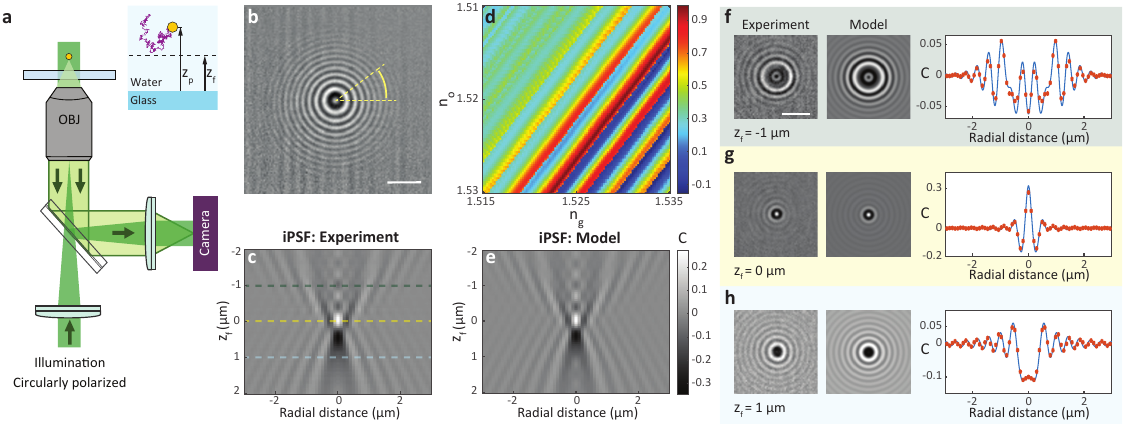}   
  \caption{(a) Schematics of a wide-field iSCAT setup. The inset shows the close-up of the relation between the axial position of the particle ($z_p$) and the focal plane ($z_f$). (b) iSCAT image of a \SI{40}{\nano\metre} GNP placed on cover glass and immersed in water, after temporal median background correction. (c) Measured iPSF stack of the GNP in (b) at different focal plane positions averaged over the azimuthal angle. (d) Normalized correlation values (see color bar) between the measured iPSF and the modeled iPSF with different $n_g$ and $n_o$. (e) Calibrated iPSF model based on the optimized values of refractive indices of oil ($n_o = 1.51833$) and glass ($n_g = 1.52696$) and glass thickness ($t_g = \SI{170}{\micro\meter}$). (f-h) Comparison between the measured and modeled iPSF including overlays of their radial profiles for focal plane at $z_f = \SI{-1}{\micro\meter}$ (f), $z_f = \SI{0}{\micro\meter}$ (g), and $z_f = \SI{1}{\micro\meter}$ (h). The right-hand plots show the average of the radial profiles computed over all angles for each case. Red symbols represent the experimental data. The blue curves show the respective theoretical models.}
  \label{setup_PSF_calibration}   
\end{figure*}   

To establish an accurate 3D model of the iPSF for a given optical setup (see Figure \ref{setup_PSF_calibration}a, we first measured the iPSF profile of single nanoparticles at the water-glass interface as the focal plane of the microscope objective was scanned through a range of $\SI{4}{\micro\metre}$. Here, we used circularly polarized light to average over induced dipole orientations, yielding iPSFs with circular symmetry (see Figure \ref{setup_PSF_calibration}b). Thus, we averaged the radial profile of the iPSF over the azimuthal angle at each focal plane and used the outcome as a representation. An example of a radial iPSF stack from a \SI{40}{\nano\meter} GNP is depicted in Figure \ref{setup_PSF_calibration}c. 

To optimize the model, we maximized the Pearson correlation value between the experimental and modeled iPSF stacks, considering different setup parameters such as the thickness of the cover glass ($t_g$) as well as the refractive indices of the immersion oil ($n_o$) and glass ($n_g$). Figure \ref{setup_PSF_calibration}d illustrates an example of the correlation value optimization process as a function of $n_o$ and $n_g$. The diagonal trend accounts for the compensation of the accumulated phase in the glass substrate and immersion oil. Figure \ref{setup_PSF_calibration}e exhibits the corresponding modelled iPSF stack for the GNP under study, demonstrating a remarkable concurrence with 99\% correlation with the experimental measurements for $n_o = 1.51833$ and $n_g = 1.52696$. To highlight the asymmetry of the iPSF relative to the focal plane, in Figures \ref{setup_PSF_calibration}f-h we depict the experimental and modeled iPSF images along with their radial profiles for $z_f = \SI{-1}{\micro\meter}$, $z_f = \SI{0}{\micro\meter}$, and $z_f = \SI{1}{\micro\meter}$, respectively.
 
\begin{figure*}[t!]   
  \includegraphics[width=0.952\textwidth]{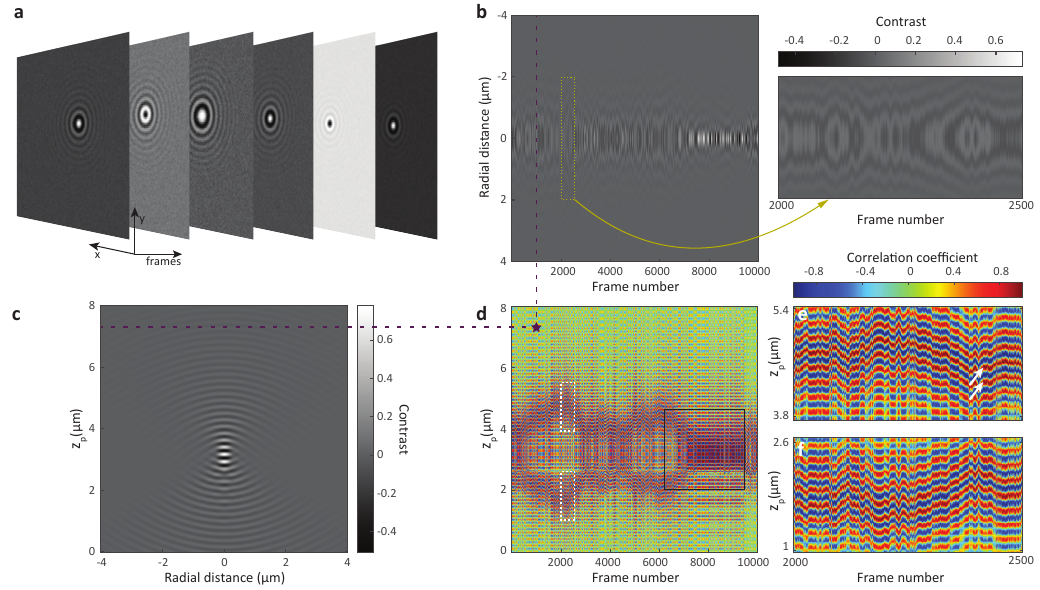}   
  \caption{(a) Synthetic iSCAT images of a \SI{40}{\nano\metre} GNP diffusing in water in 3D considering shot noise based on the camera electron well capacity. (b) Radial iPSF cross section of the GNP throughout its trajectory. The inset shows a close-up view of the region marked by the yellow rectangle. (c) Calculated iPSF model of a \SI{40}{\nano\metre} GNP for the focal plane set at \SI{3.2}{\micro\metre} above the cover glass. (d) Correlation map of the trajectory. The star indicates the correlation coefficient between the experimental and model iPSF profiles indicated by dashed lines in (b) and (c). The black square marks the region where the stripes from both sides of the focal plane join as $z_p$ approaches $z_f$. (e) Close-up view of the correlation map within the upper white dashed rectangle in (d), highlighting the striped patterns above the focal plane ($z_p > z_f$). (f) A close-up of a similar section (lower dashed rectangle in (d)) shows the details beneath the focal plane ($z_p < z_f$), revealing the partial mirror symmetry of the stripes in relation to the focal plane. (d-f) share the same color bar, representing normalized correlation values.} 
  \label{cross_section_correlationmap}   
\end{figure*}

Next, we computed synthetic trajectories for a GNP that experienced Brownian motion in water. The random step sizes of such a particle follow a normal distribution with a standard deviation of $\sigma = \sqrt{2D\Delta t}$, where $D$ is the particle's diffusion coefficient in the medium, and $\Delta t$ is the time between two consecutive frames. We reconstructed images at a frame rate of 100\,kHz to match our experimental acquisition rate acquired by a high-speed camera (Phantom V1610) with a full-well capacity of 23,200 electrons (see Figure \ref{cross_section_correlationmap}a).   
   
To localize a particle in the lateral plane, we applied radial variance transform (RVT) \cite{Kashkanova-2021} to the iPSF in each frame. This information was used to generate a cross-section map (see Figure \ref{cross_section_correlationmap}b), representing the evolution of the radial iPSF profile over time. We then employed a normalized correlation map in order to compare the temporal evolution of the experimental radial profiles to those of the model. The normalized correlation map is calculated as
\begin{equation}
\resizebox{0.89\linewidth}{!}{
  $\begin{aligned}
    \rho(i,z_p) = \frac{1}{N-1} & \sum\limits_{r=r_{\scaleto{1\mathstrut}{4pt}}}^{{r_{\scaleto{N\mathstrut}{4pt}}}} \Bigg( \biggl(\frac{RP_e(i,r) - \langle RP_e(i,r) \rangle}{\sqrt{\langle RP_e(i,r)^2 \rangle-\langle RP_e(i,r) \rangle^2}}\biggr) \\
    & \times \biggl( \frac{RP_m(z_p,r) - \langle RP_m(z_p,r) \rangle}{\sqrt{\langle RP_m(z_p,r)^2 \rangle-\langle RP_m(z_p,r) \rangle^2}}\biggr) \Biggr),
  \end{aligned}$
}
\label{eq:3}
\end{equation}
where radial profiles $RP_e(i,r)$ and $RP_m(z_p,r)$ represent the radial cross sections of the experimental and modelled iPSF, respectively. $RP(i,r)$ is a function of the frame index ($i$) and the discretized radial distance from the center of the iPSF to a given pixel ($r$). The parameter $z_p$ denotes the axial position of the particle, and $\langle \cdot \rangle$ represents averaging over $r$. The dashed lines in Figure\,\ref{cross_section_correlationmap}b,c along with the star in Figure\,\ref{cross_section_correlationmap}d exemplify the process, in which the radial profile from each frame is compared to the model at various $z_p$. 

Figure \ref{cross_section_correlationmap}d shows the resulting correlation map $\rho(i,z_p)$. Figure \ref{cross_section_correlationmap}e,f displays a close-up of two regions marked in (d). The phase difference caused by the extra travel between the cover glass and the particle leads to the modulation of the correlation map along $z_p$ so that the normalized correlation value at a particular frame oscillates rapidly with a periodicity of $\frac{\lambda}{2}$. This results in the fluctuation of the correlation values between -1 and 1 along the vertical axis, giving rise to a striped pattern. If the frame rate is sufficiently high such that the particle's axial displacement between two consecutive frames does not exceed $\frac{\lambda}{4}$, these areas will be linked to one of their neighboring frames on the correlation map. 

Before elaborating on the algorithm, we remark that a direct assignment of the axial location to the highest correlation values in a frame-by-frame procedure may lead to erroneous results, as various noise factors and setup imperfections can give rise to three potential scenarios: 1) The maximum correlation value within a given frame might not be decipherable among different stripes in the presence of noise. The white arrows in Figure \ref{cross_section_correlationmap}e  show an example of two very close correlation values of 0.99 and 0.98 in two neighboring stripes. 2) The extracted axial position of the particle undergoes jumps between the stripes situated above and below the focal plane (shown in Figures\,\ref{cross_section_correlationmap}e, and \ref{cross_section_correlationmap}f). 3) When the particle is near the focal plane the iPSF exhibits the least spatial features, as the scattered light is mostly concentrated in the center of the iPSF. Hence, the difference between the correlation value corresponding to the true $z_p$ and its respective mirror on the other side of the focal plane becomes minimal. Consequently, using the maximum correlation at each frame yields inaccurate localization. Furthermore, as highlighted in the dark square in Figure\,\ref{cross_section_correlationmap}d, axial tracking becomes even more challenging when a particle repeatedly crosses the focal plane along its trajectory. To overcome these challenges, we implement an algorithm that leverages the full spatio-temporal properties of the correlation map, thus, using the information of all the frames for determining the particle's axial location. Our algorithm also utilizes a graph representation of the correlation map to determine the axial position when the particle diffuses near the focal plane.

\begin{figure*}[t!]   
  \includegraphics[width=\textwidth]{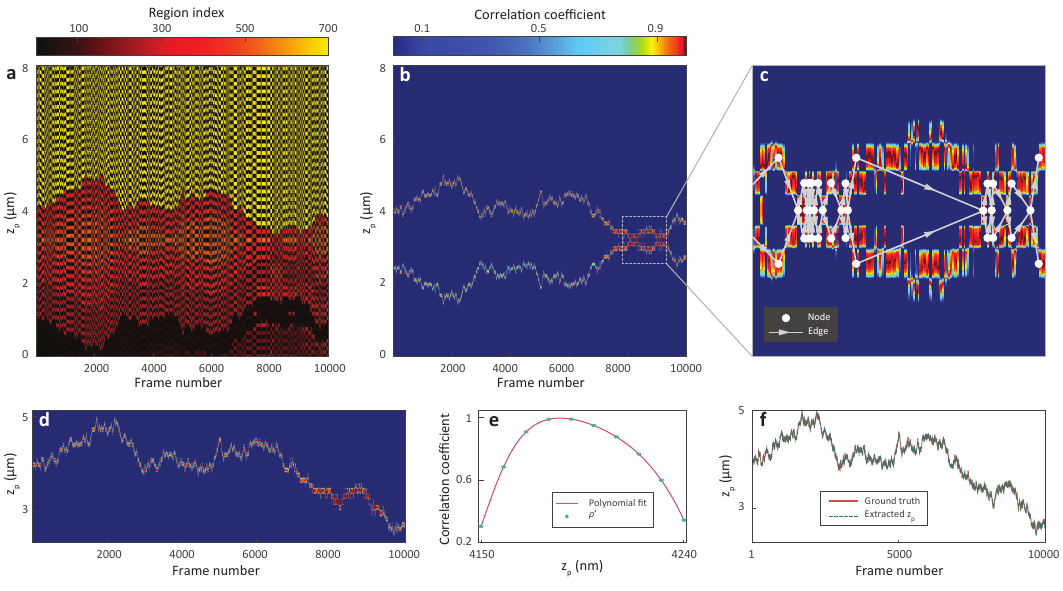}   
  \caption{(a) Labeled correlation map. Color bar shows the integer numbers assigned to each region. (b) The selected region on the correlation map with the highest correlation score. Color bar represents the correlation coefficients. (c) A magnified view of a selected region marked by the white dashed box in (b), where the particle crosses the focal plane multiple times. Lines and circles indicate the edges and nodes of the branching graph. (d) Debranched selected region of (b). (e) Example of adaptive polynomial fitting for finding the location of the maximum correlation value for a frame. (f) Comparison of the extracted axial position with the ground truth.}
  \label{correlationmap_debranching_graph}   
\end{figure*}

We start by finding the region with the highest total correlation value. To achieve that, we first convert the correlation map to a binary image by setting a global threshold at zero. Next, we segment and isolate connected regions within this binary map that have values of 1. This results in various regions $R_j$, which we label with unique integer numbers. Figure \ref{correlationmap_debranching_graph}a illustrates more than 900 regions that arise from the correlation map in Figure \ref{cross_section_correlationmap}d. Each region $R_j$ is then assigned a score $S_{R_j}$ calculated as the sum of the maximum correlation coefficients over all its frames,
\begin{equation}   
  S_{R_j} = \sum_i \max_{z_p\in R_j}(\rho(i,z_p)).   
   \label{eq:4}   
\end{equation} 
The region with the highest score, denoted as $R_{max}$, is selected for further analysis. Then the binary mask of $R_{max}$ is multiplied with the original correlation map to obtain a new map that exclusively contains the values corresponding to $R_{max}$. Figure \ref{correlationmap_debranching_graph}b shows that this procedure results in a clear selection of a region ($\rho^\prime(i,z_p)$) with high correlation values. The two branches within the beginning of the selected region above and below the focal planes can be avoided if one adjusts the focal plane to the cover glass interface or sets it well above the tracking region. We choose to place the focal plane roughly in the middle of the volume of interest because this increases the tracking range.
   
Branching within the selected region in Figure \ref{correlationmap_debranching_graph}b prevents one from determining a maximum correlation value for the axial localization. Moreover, if the local maximum values associated with two or more branches within the selected region are close to each other, frame-wise assignment of the maximum correlation can experience false jumps in the 3D trajectory, leading to inaccurate localization. To overcome this hurdle, we identify the branching points (the frames at which the number of branches varies) and create a directional graph to represent these and the sections in $\rho^\prime(i,z_p)$. Figure \ref{correlationmap_debranching_graph}c depicts the correlation map near the focus and its corresponding graph representation, illustrating occurrences of multiple branching (or complex branching) within the selected region as the particle diffuses near the focal plane. 

We divide the selected region into sections in which the number of branches remains constant. The nodes and edges of the graph represent branching points and sections in $\rho^\prime(i,z_p)$, respectively. The directionality of the edges in the graph signifies the chronological order of the frames in the trajectory, ensuring that the paths progress forward in time. An example of the directional graph is presented in Figure \ref{correlationmap_debranching_graph}c. Every edge of the graph has a distance
\begin{equation}   
  b_m = \frac{1}{\sum_{i\in B_m} \max_{z_p\in B_m}(\rho^\prime (i,z_p))} ,   
   \label{eq:5}   
\end{equation}     
where $m$ is the index of the branch $B_m$. Distances $b_m$ calculated by Eq.\,\ref{eq:5} are inversely related to the correlation values. Thus, the path with the minimum distance in the graph includes the largest sum of the correlation values along the trajectory. For a given source node, Dijkstra's algorithm \cite{Dijkstra-1959} can find the shortest path between any two nodes in the graph with an optimized computational overhead. Following this procedure, the final single-branch correlation map is reconstructed from the shortest path of the graph, which is depicted in Figure \ref{correlationmap_debranching_graph}d. We remark that computing the large number of possible combinations would present a daunting challenge. For instance, if the selected region toggles 23 times between one and two branches, the number of alternative paths from the first to the last frame amounts to $2^{23}$ ($\approx 8\times 10^6$). A brute-force approach to determining the optimal path is, thus, not viable for long trajectories. 

To refine the axial localization beyond the axial discretization of the modelled iPSF, we fit an adaptive polynomial function to the correlation values at each frame, whereby the degree of the polynomial depends on the sampling rate of the iPSF along $z_p$ (\SI{1}{\nano\metre} in this example; see Figure \ref{correlationmap_debranching_graph}e). The maximum of the fitted polynomial allows us to extract the particle's axial position along the trajectory. As seen in Figure \ref{correlationmap_debranching_graph}f, the algorithm accurately localizes the axial position when compared to the ground truth. In a representative example, using a standard desktop computer with a hexa-core processor and 64GB RAM, a trajectory of 10,000 frames with 48 nodes and 73 branches (\SI{3.35e7}{} possible outcomes) was processed in approximately 25 seconds. This demonstrates the practicality of our method for real-world applications without excessive processing time.

\subsection{Localization error and robustness}   

We now assess our algorithm's performance under various conditions. The axial localization error ($\delta z_p$) depends on $z_f$, $z_p$, shot noise, and lateral localization error ($\delta r_p$). The geometry of the iPSF is influenced by both $z_f$ and $z_p$, which in turn impacts $\delta z_p$. Shot noise and lateral localization precision both affect the extracted radial profile, the former by modifying the average over azimuthal angles and the latter by introducing an error in the identification of the iPSF's center of symmetry. 

\begin{figure*}[t!]   
  \includegraphics[width=\textwidth]{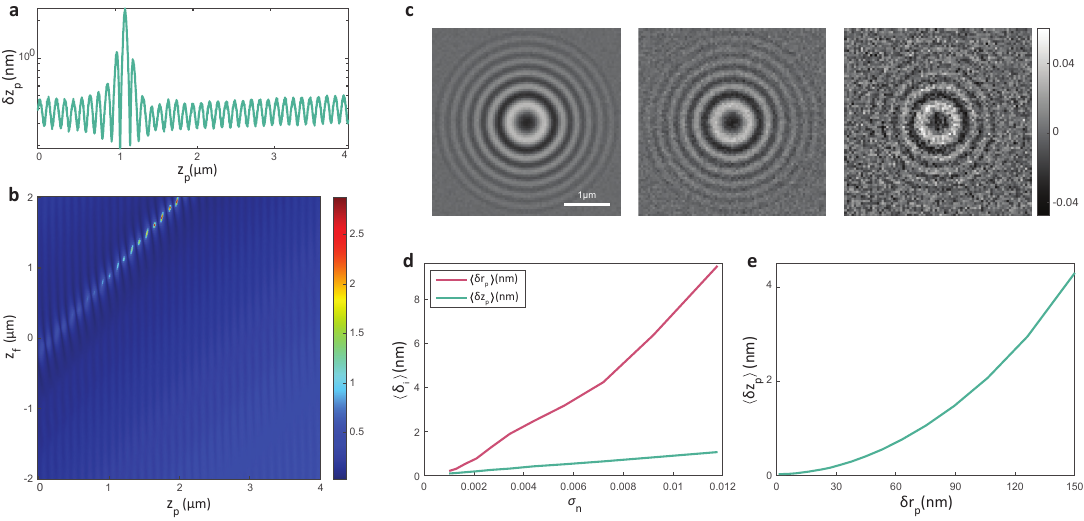}   
  \caption{ (a) Axial localization error as a function of the true axial position of a \SI{40}{\nano \metre} GNP for $z_f = \SI{1}{\micro \metre}$. (b) Standard deviation of the error ($\sigma_z(\SI{}{\nano\meter})$) in estimating the axial position of a \SI{40}{\nano \metre} GNP due to the presence of noise for different focal planes and particle axial distances. In (a,b) the normalized shot noise level is $\sigma_n=\SI{6.56e-3}{}$. (c) iSCAT images of a \SI{40}{\nano \metre} GNP in water at \SI{5.5}{\micro \metre} above the cover glass with focal plane set at \SI{3.2}{\micro \metre}. From left to right: the normalized standard deviation of shot noise is \SI {1e-3}, \SI {3.43e-3}, and \SI {1.18e-2}, respectively. The color bar represents the iSCAT contrast. (d) Lateral and axial localization uncertainty as a function of shot noise standard deviation for a \SI{40}{\nano \metre} GNP in water. The uncertainty is averaged over \SI{8}{\micro \metre} axial range with focal plane at \SI{3.2}{\micro \metre} above the cover glass. (e) Axial localization uncertainty as a function of the lateral position error for the same axial range and focal plane position as in (d), when $\sigma_n=\SI{0}{}$.}   
  \label{CRLB_robustness}   
\end{figure*}

First, we assess how $z_f$ and $z_p$ influence $\delta z_p$. Here, we assume a particle is laterally fixed (i.e., $\delta r_p = 0$) and consider a fixed level of shot noise, anticipated from our measurements. For every $z_f$ and $z_p$, we generate a large number (1000) of random realizations of the shot noise and apply our localization algorithm to the resulting noisy iPSFs to localize the particle. The axial localization error for every $z_p$ and for a given $z_f$ is then calculated by measuring the standard deviation of the differences between the retrieved positions and their known axial locations. Figure \ref{CRLB_robustness}a shows the resulting axial localization error for a \SI{40}{\nano\metre} GNP as a function of $z_p$. The axial range spans \lbrack 0, 4\rbrack\,\SI{}{\micro\metre} when $z_f = \SI{1}{\micro\metre}$ above the cover glass interface. This procedure is repeated for a series of focal planes in the range $z_f = \lbrack -2, 2 \rbrack\,\SI{}{\micro\metre}$. Figure \ref{CRLB_robustness}b displays the resulting $\sigma_z$ as a function of $z_f$ and $z_p$. As the particle approaches the focal plane, its iPSF exhibits fewer radial features, leading to an increase in axial uncertainty. Nevertheless, the highest error in estimating the axial position remains only a few nanometers. The mean localization error within $ z_p = \lbrack 0, 4 \rbrack\,\SI{}{\micro\metre}$ and $z_f = \lbrack -2, 2 \rbrack\,\SI{}{\micro\metre}$ amounts to \SI{0.2}{\nano\metre}.
   
We also evaluated the impact of the shot noise on $\delta z_p$ and $\delta r_p$ by generating iSCAT images of a \SI{40}{\nano\metre} GNP in water. Figure \ref{CRLB_robustness}c illustrates the iPSF at different shot-noise levels. We simulated videos for a particle undergoing linear axial movement in the range $z_p= \lbrack 0, 8 \rbrack\,\SI{}{\micro\metre}$ with $z_f= \SI{3.2}{\micro\metre}$ above the cover glass. We then applied our 3D tracking algorithm, whereby $\delta z_p$ and $\delta r_p$ were averaged over $z_p$ for each noise level, ranging from $\sigma_n=\SI{1e-3}{}\,\text{to}\, \SI{1.2e-2}{}$. Here, we define the normalized shot noise level as $\sigma_n = \frac{1}{\sqrt{N_e}}$ with $N_e$ representing the average number of electrons per camera pixel. As shown in Figure \ref{CRLB_robustness}d, our axial localization algorithm achieves higher precision compared to the lateral localization using the state-of-the-art RVT method. This is in agreement with the predictions of a Cramér–Rao lower bound analysis for localization in the axial direction in iSCAT microscopy \cite{Dong-2021}.

As previously stated, our algorithm operates on the radial profiles that are obtained after performing the lateral localization. To examine the sensitivity of the algorithm to the lateral localization error, we set a range of offsets for the lateral localization in the interval $\lbrack 0, 150 \rbrack$\,nm, which is close to the diffraction limit in our setup. In this analysis, we did not add shot noise to the data. In Figure \ref{CRLB_robustness}e, we present the average axial localization error $\langle \delta z_p \rangle$ over the range $z_p = \lbrack 0, 8 \rbrack\,\SI{}{\micro\metre}$ as a function of the lateral offset. This average error is calculated by first determining the absolute error in the axial localization at each $z_p$ and then computing the mean of these absolute errors across different axial positions. We find that $\langle \delta z_p \rangle$ remains below \SI{5}{\nano\metre} even with a lateral localization offset of \SI{150}{\nano\metre}.  

When considering stationary particles, an extended integration time allows for more photon collection, which reduces the effect of shot noise and enhances the signal-to-noise ratio. Consequently, as depicted in Figure \ref{CRLB_robustness}d, longer integration times (lower $\sigma_n$) provide higher lateral and axial localization precisions. However, when dealing with moving nanoparticles, the integration time must be chosen carefully to avoid motion blurring.

\subsection{Experimental results}   
     
\begin{figure*}[h!]   
  \includegraphics[width=\textwidth]{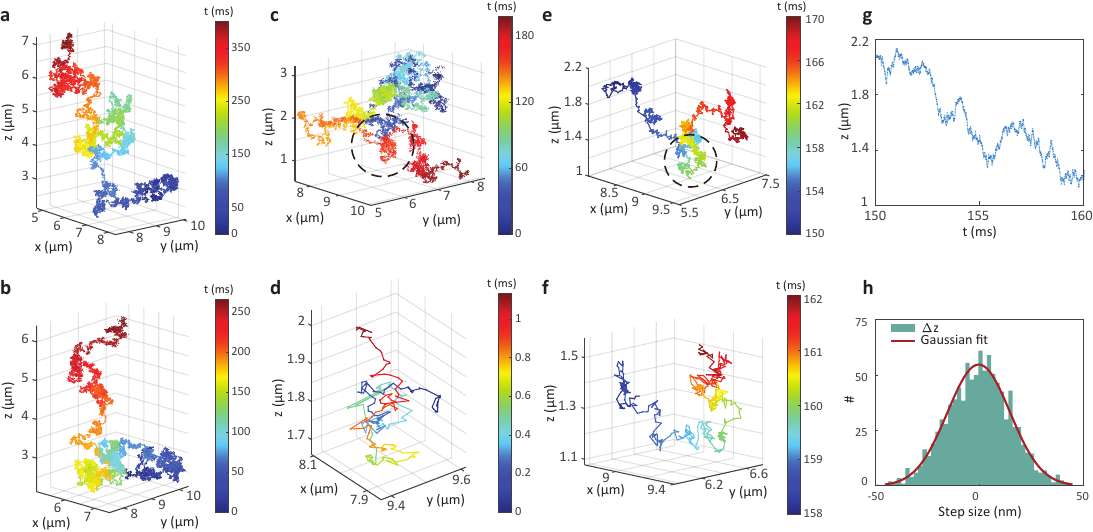}   
  \caption{ Exemplary 3D trajectories of GNPs with diameters of \SI{80}{\nano\metre} (a), \SI{60}{\nano\metre} (b), \SI{40}{\nano\metre} (c), and \SI{10}{\nano\metre} (d) diffusing in water. Trajectories that were shorter than 1000 frames were excluded from the analysis. For \SI{10}{\nano\meter} GNPs, however, we chose a lower threshold of 200 frames. Color bars represent time. (e) A close-up of the trajectory shown in (c). (f) An enlarged view of the marked region in (e). (g) Axial position as a function of time for part of the trajectory in (e) . (h) Distribution of axial step sizes in (g) fitted with a Gaussian function. }   
  \label{threeD_trajectories}   
\end{figure*}   
We now present an experimental demonstration of 3D tracking applied to GNPs of different sizes diffusing in water. We used a cover glass (Schott D 263) with a thickness of \SI{170}{\micro\metre} and created a liquid chamber by placing a gasket on it (CoverWell\textsuperscript{TM}). Then we added \SI{110}{\micro\litre} of DI water and \SI{7}{\micro\litre} of a suspension containing GNPs. In order to ensure mechanical stability, the sample rested for approximately 15 minutes. To control and assess the position of the focal plane, we marked the upper surface of the cover glass with spin coated GNPs (diameter \SI{80}{\nano\metre}) prior to the chamber assembly. The axial position of the cover glass was calibrated relative to the maximum-bright central contrast of the \SI{80}{\nano\metre} GNPs. After localizing the cover glass surface, we used the calibration of the piezo-electric scanner to displace the sample by a precise amount.

Videos of \SI{60}{\nano\metre}, \SI{40}{\nano\metre} and \SI{30}{\nano\metre} GNPs were recorded at a frame rate of \SI{100}{\kilo\hertz}. For \SI{80}{\nano\metre} and \SI{10}{\nano\metre} GNPs, we used \SI{70}{\kilo\hertz} and \SI{200}{\kilo\hertz} frame rates, respectively. To compensate for laser power fluctuations, we normalized the pixel values of each frame by dividing them by the sum of the pixel values of that frame. Then the background was subtracted using temporal median background correction, and RVT was applied to extract radial profiles in each frame. In the final step, the axial localization algorithm was applied. Figure \ref{threeD_trajectories}a-d shows examples of 3D trajectories from freely diffusing GNPs of different sizes.

Figure\,\ref{threeD_trajectories}e provides a close-up view of the trajectory within the dotted circle marked in Figures \ref{threeD_trajectories}c, and Figure\,\ref{threeD_trajectories}f displays a further close-up view of the trajectory within the dotted circle marked in Figures \ref{threeD_trajectories}e. Figure \ref{threeD_trajectories}g shows the temporal dependence of $z_p$ over 10\,ms of the trajectory in Figure \ref{threeD_trajectories}e, revealing small axial displacements in sub-ms time scale. The axial step size distribution in Figure \ref{threeD_trajectories}h reveals a histogram close to a Gaussian function as expected from a Brownian motion. By fitting a line to the mean square displacement plots of the 3D trajectories, we extract the mean value of the diffusion coefficients ($\langle D_{3D} \rangle$). For 10\,nm GNPs in water at \SI{297}{\kelvin}, we obtain $\langle D_{3D} \rangle = \SI{43.7}{} \pm \SI{1}{{{\micro\meter}^2}/{\second}}$, which is in very good agreement with the theoretical prediction of $D = \SI{44.1}{{{\micro\meter}^2}/{\second}}$. To the best of our knowledge, our work presents an unprecedented spatio-temporal precision in 3D tracking of small nanoparticles in highly diffusive liquids, such as water. 

\section{Conclusions and outlook}   
   
In most single-particle tracking applications, it is highly advantageous to use very small probes in order to avoid perturbation of the native phenomena under study due to the finite size of the probe. Over the years, particle sizes ranging from several 100\,nm down to single quantum dots or molecules have been used, albeit with varying levels of performance in terms of localization precision, accuracy and speed. In this work, we presented an experimental and algorithmic pipeline for high-precision 3D tracking of individual nanoparticles over a large range of tens of micrometers in the axial direction and at a temporal resolution as high as \SI{5}{\micro\second}. Our approach exploits the full information in the iSCAT point-spread function to establish a correlation between the experimental and model radial profiles in each frame of a video. By employing graph theory and Dijkstra's algorithm, we differentiate the particle's position above and below the focus in the temporal map of the correlation coefficients. We demonstrated an unprecedented precision of \SI{0.2}{\nano\meter} for a \SI{40}{\nano\meter} GNP as validated by simulations of synthetically generated noisy iSCAT images. We have also successfully applied our algorithm to experimental data, extracting the 3D positions of diffusing nanoparticles with sizes ranging from \SI{10}{\nano\meter} to \SI{80}{\nano\meter}. The superior axial localization precision of our algorithm, with an error that is smaller than the lateral localization error, allows us to assess the diffusion coefficient with greater precision. 3D SPT of very small nanoparticles holds great promise in studies of diffusion and transport phenomena in many areas of science and technology. 
         
\section*{Acknowledgement}   
We thank Anna Kashkanova, Mohammad Musavinezhad, Alexey Shkarin, and Reza Gholami Mahmoodabadi for helpful discussions. We are grateful to Anna Kashkanova for the careful reading of the manuscript and insightful comments. We thank the Max Planck Society for financial support.

   
   
   
\bibliography{references}

\providecommand{\latin}[1]{#1}
\makeatletter
\providecommand{\doi}
  {\begingroup\let\do\@makeother\dospecials
  \catcode`\{=1 \catcode`\}=2 \doi@aux}
\providecommand{\doi@aux}[1]{\endgroup\texttt{#1}}
\makeatother
\providecommand*\mcitethebibliography{\thebibliography}
\csname @ifundefined\endcsname{endmcitethebibliography}  {\let\endmcitethebibliography\endthebibliography}{}
\begin{mcitethebibliography}{34}
\providecommand*\natexlab[1]{#1}
\providecommand*\mciteSetBstSublistMode[1]{}
\providecommand*\mciteSetBstMaxWidthForm[2]{}
\providecommand*\mciteBstWouldAddEndPuncttrue
  {\def\EndOfBibitem{\unskip.}}
\providecommand*\mciteBstWouldAddEndPunctfalse
  {\let\EndOfBibitem\relax}
\providecommand*\mciteSetBstMidEndSepPunct[3]{}
\providecommand*\mciteSetBstSublistLabelBeginEnd[3]{}
\providecommand*\EndOfBibitem{}
\mciteSetBstSublistMode{f}
\mciteSetBstMaxWidthForm{subitem}{(\alph{mcitesubitemcount})}
\mciteSetBstSublistLabelBeginEnd
  {\mcitemaxwidthsubitemform\space}
  {\relax}
  {\relax}

\bibitem[Saxton and Jacobson(1997)Saxton, and Jacobson]{Saxton-1997}
Saxton,~M.~J.; Jacobson,~K. {Single-particle tracking: applications to membrane dynamics}. \emph{Annu. Rev. Biophys. Biomol. Struct.} \textbf{1997}, \emph{26}, 373--399\relax
\mciteBstWouldAddEndPuncttrue
\mciteSetBstMidEndSepPunct{\mcitedefaultmidpunct}
{\mcitedefaultendpunct}{\mcitedefaultseppunct}\relax
\EndOfBibitem
\bibitem[Manzo and Garcia-Parajo(2015)Manzo, and Garcia-Parajo]{Manzo-2015}
Manzo,~C.; Garcia-Parajo,~M.~F. {A review of progress in single particle tracking: from methods to biophysical insights}. \emph{Rep. Prog. Phys.} \textbf{2015}, \emph{78}, 124601\relax
\mciteBstWouldAddEndPuncttrue
\mciteSetBstMidEndSepPunct{\mcitedefaultmidpunct}
{\mcitedefaultendpunct}{\mcitedefaultseppunct}\relax
\EndOfBibitem
\bibitem[Nguyen \latin{et~al.}(2023)Nguyen, Chen, Chen, and Yeh]{Nguyen-2023}
Nguyen,~T.~D.; Chen,~Y.-I.; Chen,~L.~H.; Yeh,~H.-C. Recent Advances in Single-Molecule Tracking and Imaging Techniques. \emph{Annu. Rev. Anal. Chem.} \textbf{2023}, \emph{16}, 253--284\relax
\mciteBstWouldAddEndPuncttrue
\mciteSetBstMidEndSepPunct{\mcitedefaultmidpunct}
{\mcitedefaultendpunct}{\mcitedefaultseppunct}\relax
\EndOfBibitem
\bibitem[Mortensen \latin{et~al.}(2010)Mortensen, Churchman, Spudich, and Flyvbjerg]{mortensen-2010}
Mortensen,~K.~I.; Churchman,~L.~S.; Spudich,~J.~A.; Flyvbjerg,~H. Optimized localization analysis for single-molecule tracking and super-resolution microscopy. \emph{Nat. Methods} \textbf{2010}, \emph{7}, 377--381\relax
\mciteBstWouldAddEndPuncttrue
\mciteSetBstMidEndSepPunct{\mcitedefaultmidpunct}
{\mcitedefaultendpunct}{\mcitedefaultseppunct}\relax
\EndOfBibitem
\bibitem[Huhle \latin{et~al.}(2015)Huhle, Klaue, Brutzer, Daldrop, Joo, Otto, Keyser, and Seidel]{huhle-2015-tweezer}
Huhle,~A.; Klaue,~D.; Brutzer,~H.; Daldrop,~P.; Joo,~S.; Otto,~O.; Keyser,~U.~F.; Seidel,~R. Camera-based three-dimensional real-time particle tracking at kHz rates and {\AA}ngstr{\"o}m accuracy. \emph{Nat. Commun.} \textbf{2015}, \emph{6}, 5885\relax
\mciteBstWouldAddEndPuncttrue
\mciteSetBstMidEndSepPunct{\mcitedefaultmidpunct}
{\mcitedefaultendpunct}{\mcitedefaultseppunct}\relax
\EndOfBibitem
\bibitem[von Diezmann \latin{et~al.}(2017)von Diezmann, Shechtman, and Moerner]{VonDiezmann2017}
von Diezmann,~L.; Shechtman,~Y.; Moerner,~W. Three-dimensional localization of single molecules for super-resolution imaging and single-particle tracking. \emph{Chem. Rev.} \textbf{2017}, \emph{117}, 7244--7275\relax
\mciteBstWouldAddEndPuncttrue
\mciteSetBstMidEndSepPunct{\mcitedefaultmidpunct}
{\mcitedefaultendpunct}{\mcitedefaultseppunct}\relax
\EndOfBibitem
\bibitem[Reina \latin{et~al.}(2018)Reina, Galiani, Shrestha, Sezgin, de~Wit, Cole, Lagerholm, Kukura, and Eggeling]{Reina2018}
Reina,~F.; Galiani,~S.; Shrestha,~D.; Sezgin,~E.; de~Wit,~G.; Cole,~D.; Lagerholm,~B.~C.; Kukura,~P.; Eggeling,~C. Complementary studies of lipid membrane dynamics using iSCAT and super-resolved fluorescence correlation spectroscopy. \emph{J. Phys. D: Appl. Phys.} \textbf{2018}, \emph{51}, 235401\relax
\mciteBstWouldAddEndPuncttrue
\mciteSetBstMidEndSepPunct{\mcitedefaultmidpunct}
{\mcitedefaultendpunct}{\mcitedefaultseppunct}\relax
\EndOfBibitem
\bibitem[Weeks \latin{et~al.}(2000)Weeks, Crocker, Levitt, Schofield, and Weitz]{Weeks2000}
Weeks,~E.~R.; Crocker,~J.~C.; Levitt,~A.~C.; Schofield,~A.; Weitz,~D.~A. Three-dimensional direct imaging of structural relaxation near the colloidal glass transition. \emph{Science} \textbf{2000}, \emph{287}, 627--631\relax
\mciteBstWouldAddEndPuncttrue
\mciteSetBstMidEndSepPunct{\mcitedefaultmidpunct}
{\mcitedefaultendpunct}{\mcitedefaultseppunct}\relax
\EndOfBibitem
\bibitem[Metzler \latin{et~al.}(2014)Metzler, Jeon, Cherstvy, and Barkai]{Metzler2014}
Metzler,~R.; Jeon,~J.-H.; Cherstvy,~A.~G.; Barkai,~E. Anomalous diffusion models and their properties: non-stationarity, non-ergodicity, and ageing at the centenary of single particle tracking. \emph{Phys. Chem. Chem. Phys.} \textbf{2014}, \emph{16}, 24128--24164\relax
\mciteBstWouldAddEndPuncttrue
\mciteSetBstMidEndSepPunct{\mcitedefaultmidpunct}
{\mcitedefaultendpunct}{\mcitedefaultseppunct}\relax
\EndOfBibitem
\bibitem[Mazaheri \latin{et~al.}(2020)Mazaheri, Ehrig, Shkarin, Zaburdaev, and Sandoghdar]{Mazaheri-2020}
Mazaheri,~M.; Ehrig,~J.; Shkarin,~A.; Zaburdaev,~V.; Sandoghdar,~V. {Ultrahigh-Speed Imaging of Rotational Diffusion on a Lipid Bilayer}. \emph{Nano Lett.} \textbf{2020}, \emph{20}, 7213--7219\relax
\mciteBstWouldAddEndPuncttrue
\mciteSetBstMidEndSepPunct{\mcitedefaultmidpunct}
{\mcitedefaultendpunct}{\mcitedefaultseppunct}\relax
\EndOfBibitem
\bibitem[Beckwith and Yang(2021)Beckwith, and Yang]{Beckwith2021}
Beckwith,~J.~S.; Yang,~H. {Sub-millisecond Translational and Orientational Dynamics of a Freely Moving Single Nanoprobe}. \emph{J. Phys. Chem. B.} \textbf{2021}, \emph{125}, 13436--13443\relax
\mciteBstWouldAddEndPuncttrue
\mciteSetBstMidEndSepPunct{\mcitedefaultmidpunct}
{\mcitedefaultendpunct}{\mcitedefaultseppunct}\relax
\EndOfBibitem
\bibitem[Wells \latin{et~al.}(2010)Wells, Lessard, Goodwin, Phipps, Cutler, Lidke, Wilson, and Werner]{Wells-2010}
Wells,~N.~P.; Lessard,~G.~A.; Goodwin,~P.~M.; Phipps,~M.~E.; Cutler,~P.~J.; Lidke,~D.~S.; Wilson,~B.~S.; Werner,~J.~H. {Time-Resolved Three-Dimensional Molecular Tracking in Live Cells}. \emph{Nano Lett.} \textbf{2010}, \emph{10}, 4732--4737\relax
\mciteBstWouldAddEndPuncttrue
\mciteSetBstMidEndSepPunct{\mcitedefaultmidpunct}
{\mcitedefaultendpunct}{\mcitedefaultseppunct}\relax
\EndOfBibitem
\bibitem[Wang \latin{et~al.}(2019)Wang, Yi, Gdor, Hereld, and Scherer]{Wang-2019}
Wang,~X.; Yi,~H.; Gdor,~I.; Hereld,~M.; Scherer,~N.~F. {Nanoscale Resolution 3D Snapshot Particle Tracking by Multifocal Microscopy}. \emph{Nano Lett.} \textbf{2019}, \emph{19}, 6781--6787\relax
\mciteBstWouldAddEndPuncttrue
\mciteSetBstMidEndSepPunct{\mcitedefaultmidpunct}
{\mcitedefaultendpunct}{\mcitedefaultseppunct}\relax
\EndOfBibitem
\bibitem[Louis \latin{et~al.}(2020)Louis, Camacho, Bresol\'{i}-Obach, Abakumov, Vandaele, Kudo, Masuhara, Scheblykin, Hofkens, and Rocha]{Louis-2020}
Louis,~B.; Camacho,~R.; Bresol\'{i}-Obach,~R.; Abakumov,~S.; Vandaele,~J.; Kudo,~T.; Masuhara,~H.; Scheblykin,~I.~G.; Hofkens,~J.; Rocha,~S. {Fast-tracking of single emitters in large volumes with nanometer precision}. \emph{Opt. Express} \textbf{2020}, \emph{28}, 28656--28671\relax
\mciteBstWouldAddEndPuncttrue
\mciteSetBstMidEndSepPunct{\mcitedefaultmidpunct}
{\mcitedefaultendpunct}{\mcitedefaultseppunct}\relax
\EndOfBibitem
\bibitem[Thompson \latin{et~al.}(2010)Thompson, Lew, Badieirostami, and Moerner]{Moerner2010}
Thompson,~M.~A.; Lew,~M.~D.; Badieirostami,~M.; Moerner,~W. Localizing and tracking single nanoscale emitters in three dimensions with high spatiotemporal resolution using a double-helix point spread function. \emph{Nano Lett.} \textbf{2010}, \emph{10}, 211--218\relax
\mciteBstWouldAddEndPuncttrue
\mciteSetBstMidEndSepPunct{\mcitedefaultmidpunct}
{\mcitedefaultendpunct}{\mcitedefaultseppunct}\relax
\EndOfBibitem
\bibitem[Taylor \latin{et~al.}(2019)Taylor, Mahmoodabadi, Rauschenberger, Giessl, Schambony, and Sandoghdar]{Taylor-2019}
Taylor,~R.~W.; Mahmoodabadi,~R.~G.; Rauschenberger,~V.; Giessl,~A.; Schambony,~A.; Sandoghdar,~V. {Interferometric scattering microscopy reveals microsecond nanoscopic protein motion on a live cell membrane}. \emph{Nat. Photonics} \textbf{2019}, \emph{13}, 480--487\relax
\mciteBstWouldAddEndPuncttrue
\mciteSetBstMidEndSepPunct{\mcitedefaultmidpunct}
{\mcitedefaultendpunct}{\mcitedefaultseppunct}\relax
\EndOfBibitem
\bibitem[Martin \latin{et~al.}(2022)Martin, Altman, Rawat, Wang, Grier, and Manoharan]{martin-2022}
Martin,~C.; Altman,~L.~E.; Rawat,~S.; Wang,~A.; Grier,~D.~G.; Manoharan,~V.~N. In-line holographic microscopy with model-based analysis. \emph{Nat. Rev. Methods Primers} \textbf{2022}, \emph{2}, 83\relax
\mciteBstWouldAddEndPuncttrue
\mciteSetBstMidEndSepPunct{\mcitedefaultmidpunct}
{\mcitedefaultendpunct}{\mcitedefaultseppunct}\relax
\EndOfBibitem
\bibitem[Verpillat \latin{et~al.}(2011)Verpillat, Joud, Desbiolles, and Gross]{Verpillat2011}
Verpillat,~F.; Joud,~F.; Desbiolles,~P.; Gross,~M. Dark-field digital holographic microscopy for 3D-tracking of gold nanoparticles. \emph{Opt. Express} \textbf{2011}, \emph{19}, 26044--26055\relax
\mciteBstWouldAddEndPuncttrue
\mciteSetBstMidEndSepPunct{\mcitedefaultmidpunct}
{\mcitedefaultendpunct}{\mcitedefaultseppunct}\relax
\EndOfBibitem
\bibitem[Lindfors \latin{et~al.}(2004)Lindfors, Kalkbrenner, Stoller, and Sandoghdar]{Lindfors-2004}
Lindfors,~K.; Kalkbrenner,~T.; Stoller,~P.; Sandoghdar,~V. {Detection and Spectroscopy of Gold Nanoparticles Using Supercontinuum White Light Confocal Microscopy}. \emph{Phys. Rev. Lett.} \textbf{2004}, \emph{93}, 037401\relax
\mciteBstWouldAddEndPuncttrue
\mciteSetBstMidEndSepPunct{\mcitedefaultmidpunct}
{\mcitedefaultendpunct}{\mcitedefaultseppunct}\relax
\EndOfBibitem
\bibitem[Taylor and Sandoghdar(2019)Taylor, and Sandoghdar]{Taylor-2019-nano}
Taylor,~R.~W.; Sandoghdar,~V. {Interferometric Scattering Microscopy: Seeing Single Nanoparticles and Molecules via Rayleigh Scattering}. \emph{Nano Lett.} \textbf{2019}, \emph{19}, 4827--4835\relax
\mciteBstWouldAddEndPuncttrue
\mciteSetBstMidEndSepPunct{\mcitedefaultmidpunct}
{\mcitedefaultendpunct}{\mcitedefaultseppunct}\relax
\EndOfBibitem
\bibitem[Piliarik and Sandoghdar(2014)Piliarik, and Sandoghdar]{Piliarik-2014}
Piliarik,~M.; Sandoghdar,~V. {Direct optical sensing of single unlabelled proteins and super-resolution imaging of their binding sites}. \emph{Nat. Commun.} \textbf{2014}, \emph{5}, 4495\relax
\mciteBstWouldAddEndPuncttrue
\mciteSetBstMidEndSepPunct{\mcitedefaultmidpunct}
{\mcitedefaultendpunct}{\mcitedefaultseppunct}\relax
\EndOfBibitem
\bibitem[Young \latin{et~al.}(2018)Young, Hundt, Cole, Fineberg, Andrecka, Tyler, Olerinyova, Ansari, Marklund, Collier, \latin{et~al.} others]{Young-2018}
Young,~G.; Hundt,~N.; Cole,~D.; Fineberg,~A.; Andrecka,~J.; Tyler,~A.; Olerinyova,~A.; Ansari,~A.; Marklund,~E.~G.; Collier,~M.~P.; others {Quantitative mass imaging of single biological macromolecules}. \emph{Science} \textbf{2018}, \emph{360}, 423--427\relax
\mciteBstWouldAddEndPuncttrue
\mciteSetBstMidEndSepPunct{\mcitedefaultmidpunct}
{\mcitedefaultendpunct}{\mcitedefaultseppunct}\relax
\EndOfBibitem
\bibitem[Dahmardeh \latin{et~al.}(2023)Dahmardeh, Mirzaalian~Dastjerdi, Mazal, K{\"o}stler, and Sandoghdar]{Dahmardeh-2023}
Dahmardeh,~M.; Mirzaalian~Dastjerdi,~H.; Mazal,~H.; K{\"o}stler,~H.; Sandoghdar,~V. Self-supervised machine learning pushes the sensitivity limit in label-free detection of single proteins below 10 {kDa}. \emph{Nat. Methods} \textbf{2023}, \emph{20}, 442--447\relax
\mciteBstWouldAddEndPuncttrue
\mciteSetBstMidEndSepPunct{\mcitedefaultmidpunct}
{\mcitedefaultendpunct}{\mcitedefaultseppunct}\relax
\EndOfBibitem
\bibitem[Kukura \latin{et~al.}(2009)Kukura, Ewers, M{\"u}ller, Renn, Helenius, and Sandoghdar]{Kukura-2009}
Kukura,~P.; Ewers,~H.; M{\"u}ller,~C.; Renn,~A.; Helenius,~A.; Sandoghdar,~V. High-speed nanoscopic tracking of the position and orientation of a single virus. \emph{Nat. Methods} \textbf{2009}, \emph{6}, 923--927\relax
\mciteBstWouldAddEndPuncttrue
\mciteSetBstMidEndSepPunct{\mcitedefaultmidpunct}
{\mcitedefaultendpunct}{\mcitedefaultseppunct}\relax
\EndOfBibitem
\bibitem[Lin \latin{et~al.}(2014)Lin, Chang, and Hsieh]{Lin-2014}
Lin,~Y.-H.; Chang,~W.-L.; Hsieh,~C.-L. {Shot-noise limited localization of single 20 nm gold particles with nanometer spatial precision within microseconds}. \emph{Opt. Express} \textbf{2014}, \emph{22}, 9159--9170\relax
\mciteBstWouldAddEndPuncttrue
\mciteSetBstMidEndSepPunct{\mcitedefaultmidpunct}
{\mcitedefaultendpunct}{\mcitedefaultseppunct}\relax
\EndOfBibitem
\bibitem[Spindler \latin{et~al.}(2016)Spindler, Ehrig, K{\"o}nig, Nowak, Piliarik, Stein, Taylor, Garanger, Lecommandoux, Alves, and Sandoghdar]{Spindler-2016}
Spindler,~S.; Ehrig,~J.; K{\"o}nig,~K.; Nowak,~T.; Piliarik,~M.; Stein,~H.~E.; Taylor,~R.~W.; Garanger,~E.; Lecommandoux,~S.; Alves,~I.~D.; Sandoghdar,~V. {Visualization of lipids and proteins at high spatial and temporal resolution via interferometric scattering ({iSCAT}) microscopy}. \emph{J. Phys. D: Appl. Phys.} \textbf{2016}, \emph{49}, 274002\relax
\mciteBstWouldAddEndPuncttrue
\mciteSetBstMidEndSepPunct{\mcitedefaultmidpunct}
{\mcitedefaultendpunct}{\mcitedefaultseppunct}\relax
\EndOfBibitem
\bibitem[Jacobsen \latin{et~al.}(2007)Jacobsen, Klotzsch, and Sandoghdar]{jacobsen_2007}
Jacobsen,~V.; Klotzsch,~E.; Sandoghdar,~V. \emph{Interferometric detection and tracking of nanoparticles}; Elsevier: Amsterdam, 2007; Vol.~3\relax
\mciteBstWouldAddEndPuncttrue
\mciteSetBstMidEndSepPunct{\mcitedefaultmidpunct}
{\mcitedefaultendpunct}{\mcitedefaultseppunct}\relax
\EndOfBibitem
\bibitem[Krishnan \latin{et~al.}(2010)Krishnan, Mojarad, Kukura, and Sandoghdar]{Krishnan-2010}
Krishnan,~M.; Mojarad,~N.; Kukura,~P.; Sandoghdar,~V. {Geometry-induced electrostatic trapping of nanometric objects in a fluid}. \emph{Nature} \textbf{2010}, \emph{467}, 692--695\relax
\mciteBstWouldAddEndPuncttrue
\mciteSetBstMidEndSepPunct{\mcitedefaultmidpunct}
{\mcitedefaultendpunct}{\mcitedefaultseppunct}\relax
\EndOfBibitem
\bibitem[{de Wit} \latin{et~al.}(2018){de Wit}, Albrecht, Ewers, and Kukura]{deWit-2018}
{de Wit},~G.; Albrecht,~D.; Ewers,~H.; Kukura,~P. {Revealing Compartmentalized Diffusion in Living Cells with Interferometric Scattering Microscopy}. \emph{Biophys. J.} \textbf{2018}, \emph{114}, 2945--2950\relax
\mciteBstWouldAddEndPuncttrue
\mciteSetBstMidEndSepPunct{\mcitedefaultmidpunct}
{\mcitedefaultendpunct}{\mcitedefaultseppunct}\relax
\EndOfBibitem
\bibitem[Mahmoodabadi \latin{et~al.}(2020)Mahmoodabadi, Taylor, Kaller, Spindler, Mazaheri, Kasaian, and Sandoghdar]{Mahmoodabadi-2020}
Mahmoodabadi,~R.~G.; Taylor,~R.~W.; Kaller,~M.; Spindler,~S.; Mazaheri,~M.; Kasaian,~K.; Sandoghdar,~V. {Point spread function in interferometric scattering microscopy (iSCAT). Part I: aberrations in defocusing and axial localization}. \emph{Opt. Express} \textbf{2020}, \emph{28}, 25969--25988\relax
\mciteBstWouldAddEndPuncttrue
\mciteSetBstMidEndSepPunct{\mcitedefaultmidpunct}
{\mcitedefaultendpunct}{\mcitedefaultseppunct}\relax
\EndOfBibitem
\bibitem[Kashkanova \latin{et~al.}(2021)Kashkanova, Shkarin, Mahmoodabadi, Blessing, Tuna, Gemeinhardt, and Sandoghdar]{Kashkanova-2021}
Kashkanova,~A.~D.; Shkarin,~A.~B.; Mahmoodabadi,~R.~G.; Blessing,~M.; Tuna,~Y.; Gemeinhardt,~A.; Sandoghdar,~V. {Precision single-particle localization using radial variance transform}. \emph{Opt. Express} \textbf{2021}, \emph{29}, 11070--11083\relax
\mciteBstWouldAddEndPuncttrue
\mciteSetBstMidEndSepPunct{\mcitedefaultmidpunct}
{\mcitedefaultendpunct}{\mcitedefaultseppunct}\relax
\EndOfBibitem
\bibitem[Dijkstra(1959)]{Dijkstra-1959}
Dijkstra,~E.~W. {A note on two problems in connexion with graphs}. \emph{Numer. Math.} \textbf{1959}, \emph{1}, 269--271\relax
\mciteBstWouldAddEndPuncttrue
\mciteSetBstMidEndSepPunct{\mcitedefaultmidpunct}
{\mcitedefaultendpunct}{\mcitedefaultseppunct}\relax
\EndOfBibitem
\bibitem[Dong \latin{et~al.}(2021)Dong, Maestre, Conrad-Billroth, and Juffmann]{Dong-2021}
Dong,~J.; Maestre,~D.; Conrad-Billroth,~C.; Juffmann,~T. {Fundamental bounds on the precision of {iSCAT}, {COBRI} and dark-field microscopy for 3D localization and mass photometry}. \emph{J. Phys. D: Appl. Phys.} \textbf{2021}, \emph{54}, 394002\relax
\mciteBstWouldAddEndPuncttrue
\mciteSetBstMidEndSepPunct{\mcitedefaultmidpunct}
{\mcitedefaultendpunct}{\mcitedefaultseppunct}\relax
\EndOfBibitem
\end{mcitethebibliography}

\end{document}